\documentclass[conference]{IEEEtran}
\IEEEoverridecommandlockouts
\usepackage{cite}
\usepackage{amsmath,amssymb,amsfonts}
\usepackage{algorithmic}
\usepackage{graphicx}
\usepackage{lipsum}
\usepackage{textcomp}
\usepackage{xcolor}
\def\BibTeX{{\rm B\kern-.05em{\sc i\kern-.025em b}\kern-.08em
    T\kern-.1667em\lower.7ex\hbox{E}\kern-.125emX}}
\def\IEEEkeywordsname{Keywords}
\begin{document}
\title{Effect of Charge-neutrality Breaking \\ on Localized Terahertz Waves \\ in a Plate of Layered Superconductor
}
\author{\IEEEauthorblockN{Nina Kvitka}
\IEEEauthorblockA{\textit{Faculty of Physics, University of Vienna,}\\
Vienna, Austria
\\ \textit{Theoretical Physics Department,}
\\ \textit{O.Ya. Usikov Institute for} \\ \textit{Radiophysics and Electronics,}\\ 
Kharkiv, Ukraine \\
nina.kvitka@univie.ac.at}
\and
\IEEEauthorblockN{Stanislav Apostolov}
\IEEEauthorblockA{\textit{Theoretical Physics Department,} 
\\ \textit{O.Ya. Usikov Institute for} \\ \textit{Radiophysics and Electronics;} \\
\textit{Theoretical Physics Department,} 
\\ \textit{V.N.Karazin Kharkiv National University,}\\ 
Kharkiv, Ukraine \\
{ORCID: 0000-0003-1348-8204}}\\
\and
\IEEEauthorblockN{Valery Yampol’skii}
\IEEEauthorblockA{\textit{Theoretical Physics Department,} 
\\ \textit{O.Ya. Usikov Institute for} \\ \textit{ Radiophysics and Electronics;}\\
\textit{Theoretical Physics Department,} 
\\ \textit{V.N.Karazin Kharkiv National University,}\\ 
Kharkiv, Ukraine \\
yam@ire.kharkov.ua
}}
\maketitle
\begin{abstract}
We study analytically the effect of charge-neutrality breaking on the spectrum of localized Josephson plasma waves (JPWs) in a plate of layered superconductor of finite thickness for the case of layers parallel to the interface. This effect results in the appearance of additional eigenmodes with anomalous dispersion and an additional forbidden region in the spectrum of localized modes near the Josephson plasma frequency. We consider both the surface and waveguide eigenmodes with symmetric and anti-symmetric field distribution with respect to the middle plane of the superconducting plate. 
\end{abstract}
\vspace{10pt} 


\section{Introduction}

Layered superconductors, e.g. high-temperature cuprate superconductors or artificially grown layered structures of 2-dimensional Josephson junctions, draw the attention of researchers because of their unique properties related to the high anisotropy. The characteristic frequency range for the electromagnetic excitations in these materials --- Josephson plasma waves (JPWs) --- is the terahertz (THz) range. Devices operating within this range have many possible practical applications in many different fields from quality control to astrophysics \cite{thz}. Another peculiar property of layered superconductors is the anomalous dispersion of JPWs that can be observed when the superconducting layers are perpendicular to the sample interfaces \cite{anom}. Anomalous dispersion results in various phenomena, e.g. negative refractive index \cite{neg} or specific form of resonant transmission \cite{anom}.

In this paper, we will investigate the eigenwaves in a finite plate of layered superconductor considering the effect of charge-neutrality breaking in geometry when the layers are parallel to the interfaces. The spectrum of these waves (both surface, decaying exponentially inside the plate, and wave-guide, oscillating across the layers) was studied in~\cite{plate} for the charge-neutral case, and the dispersion of the waves was normal in this geometry. In Ref.~\cite{en}, the surface JPWs for an infinite slab were analyzed keeping the term that is responsible for the neutrality breaking effect, and it was shown that the dispersion curve consists of two branches. The latter results in many interesting optical phenomena such as, for example, sharp absorption peaks \cite{en} or peaks in transmissivity \cite{bul}. Since the surface wave rapidly decays inside the sample, the spectrum for the surface mode, below Josephson frequency, is similar to the case of a finite plate, therefore, we will focus, first of all, on the wave-guide waves. 

\section{Model}

\begin{figure}[t]
\centerline{\includegraphics[width=0.41\textwidth]{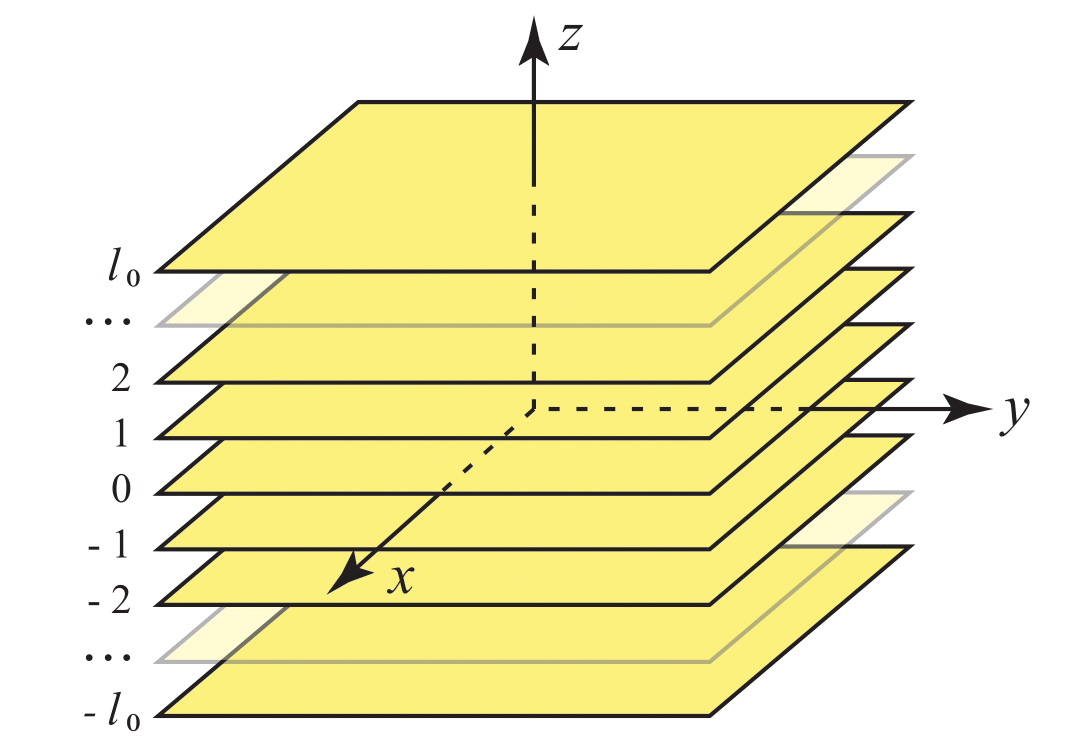}}
\caption{ The layered superconductor plate with~$(2 l_0+1)$ layers in the vacuum environment. The geometry of the problem.}
\label{fig1}
\end{figure}

We study the TM-polarized waves of the frequency~$\omega$, 
\begin{equation}
 \mathbf{E}=\{E_x,0,E_z\},\quad
 \mathbf{H}=\{0,H_y,0\},
\end{equation}
localized on a plate of layered superconductor containing $(2 l_0+1)$ thin superconducting layers, see Fig.~\ref{fig1}. We assume the layers to be parallel to the plate interfaces which are superconducting also, then the thickness of the plate is $2l_0D$, where $D$ is the period of the layered structure. 
%which is considered to be much greater than the thickness of the superconducting layer. (избыточная инфа - уже было сказано, что thin superconducting layers, и предложение перегружено)
Note that the odd number of layers is not essential for the problem and it was chosen only for reasons of simplicity of the equation below.
%simplicity of boundary conditions.

We study the electromagnetic eigenwaves with both symmetrical and anti-symmetrical magnetic field distribution with respect to the middle plane, $z=0$, of the plate.
In both cases, it suffices to determine the field in the upper half-space, $z>0$. The field in the lower half-space, $z<0$, is easily found using the mentioned symmetry.

\subsection{Electromagnetic Field in Vacuum}

In the vacuum region, $|z| > l_0 D$, we consider the plane waves to be evanescent in the $z$-direction,
\begin{equation}
   E_x,E_z,H_y\propto \exp(-i\omega t +i k_x x-k_v |z|),
   \nonumber
\end{equation}
with the well-known dispersion relation, 
\begin{equation}
   k_v = \sqrt{k_x^2-\omega^2/c^2}.
   \nonumber
\end{equation}
Here $k_x$ is the longitudinal wave number, and $c$ is the speed of light.

The ratio of tangential components of electric and magnetic fields right above the plate surface 
\begin{equation}
\label{ratio}
   \frac{E_x}{H_y}  = \frac{i c k_v}{\omega},
\end{equation}
is easily obtained from the Maxwell equations. 

\subsection{Electromagnetic Field in Superconductor}

Inside the plate, for $|z| < l_0 D$, the field distribution is determined by the set of coupled nonlinear sine-Gordon equations for the phase difference of the order parameter in the neighboring superconducting layers. 
For the JPWs, we can linearize these equations and rewrite them for the magnetic fields between the superconducting layers~\cite{en}, $H_{l}\equiv H_{y}|_{z=lD}$,
\begin{equation}
\label{sg}
   \left(1-\frac{\lambda_{ab}^2}{D^2}\partial_l^2\right)\left(\dfrac{1}{\omega_J^2}\frac{\partial^2 H_{l}}{\partial t^2}+ H_{l}-\alpha \partial_l^2 H_{l}\right)=\lambda_{c}^2\frac{\partial^2 H_{l}}{\partial x^2}.
\end{equation}
Here the operator of the discrete second derivative, $\partial_l^2$, is 
\begin{equation}
\label{oper}
\partial_l^2F_l=F_{l+1}+F_{l-1}-2 F_{l},
\end{equation}
$\lambda_{ab}$ and $\lambda_{c}$ are the London penetration depths in directions perpendicular and parallel to the layers, respectively, $\omega_J$ is the Josephson plasma frequency, $\alpha$ is the constant characterizing the effect of charge-neutrality breaking. For the natural crystals (for example, Bi2212 or Tl2212), $\alpha\sim 0.05-0.1$ \cite{bul}.

Due to the symmetry of the problem, inside the plate, the JPWs can be either symmetrical or anti-symmetrical (``$+$'' or ``$-$'' in the upper index, respectively):
\begin{align}
H_l^\pm = &C_{\pm} \exp(-i\omega t +i k_x x)
\notag
\\
\label{Hy}
&\times  \left[\exp(-k_s (l_0-l) D) \pm \exp(-k_s (l_0+l) D)\right].
\end{align}
Substituting Eq.~\eqref{Hy} into Eq.~\eqref{sg}, we get the implicit equation for the wave-number $k_s$ inside the sample:
\begin{align}
  \left(\frac{\omega^2}{\omega_J^2}-1+4 \alpha \kappa \right)   \left(1- \frac{4\lambda_{ab}^2}{D^2}\kappa\right)\label{ks}=\frac{c^2 k_x^2}{\omega_J^2\varepsilon}.
\end{align}
The last equation is quadratic with respect to $\kappa$:
\begin{equation}
\label{kappa}
    \kappa={\rm sinh}^2 \left(\frac{k_s D}{2}\right),
\end{equation}
and, therefore, we have two solutions for $k_s^2$. In the non-dissipative model, the wave number $k_s$ can be either real or imaginary, resulting in a real value of ${\rm sinh}^2(k_s D/2)$. So, the discriminant of Eq.~\eqref{ks} must be non-negative, which imposes an additional condition on the problem parameters,
%\lambda_c=c/(\omega_J \sqrt{\varepsilon})
\begin{equation}
\label{cond}
    \frac{|\alpha (D/\lambda_{ab})^2+(\omega/\omega_J)^2-1|}{2  k_s \lambda_c (D/\lambda_{ab})\sqrt{\alpha}}\geq 1.
\end{equation}
It is easy to see that, at any finite value of $\omega\not=\omega_J$, the left part of the last inequality tends to infinity at $\alpha\to 0$. So, without the charge-neutrality breaking, this condition is satisfied. 

Using the London equation for the magnetic field near the upper interface ($z=l_0 D$) in discrete approximation, 
\begin{align}
\label{Hlond}
   \frac{H_{l_0}-H_{l_0-1}}{D}\approx \frac{-i c E_{l_0}}{\lambda_{ab}^2 \omega},
\end{align}
where $E_{l}\equiv E_x{\big|}_{z=l D}$,
and expression~\eqref{Hy},
we obtain the following ratios for the tangential components of the electric and magnetic fields at $z=l_0 D$ right below the plate surface:
\begin{align}
\label{ratios}
  \frac{E_{l_0}^+}{H_{l_0}^+}  = \frac{i \omega \lambda_{ab}^2}{c D} \left(1-\frac{{\rm cosh}\left[k_s(l_0-1)D\right]}{{\rm cosh}[k_s l_0 D]}\right),\\
  \label{ratioa}
  \frac{E_{l_0}^-}{H_{l_0}^-}   = \frac{i \omega \lambda_{ab}^2}{c D} \left(1-\frac{{\rm sinh}\left[k_s(l_0-1)D\right]}{{\rm sinh}[k_s l_0 D]}\right).
\end{align}
The ratios~\eqref{ratios} and~\eqref{ratioa} are for the symmetrical and anti-symmetrical modes, respectively.  

It is easy to see that, in case of an infinitely thick plate, $l_0\to\infty$, and for $k_s^2>0$, both Eqs.~\eqref{ratios} and~\eqref{ratioa} give the same limiting expression,
\begin{equation}
\lim_{l_0\to\infty}\frac{E_{l_0}^\pm}{H_{l_0}^\pm}=\frac{i \omega \lambda_{ab}^2}{c D}\left(1-e^{-k_s D} \right),
\end{equation}
that is consistent with the result for the semi-infinite plate obtained in~\cite{en}. 
In the continuum limit, $k_s D \ll 1$, we have
\begin{eqnarray}
\label{conts}
\frac{E_{l_0}^+}{H_{l_0}^+}= \frac{i \omega \lambda_{ab}^2  k_s}{c} {\rm tanh} (k_s l_0 D),\\
\label{conta}
\frac{E_{l_0}^-}{H_{l_0}^-} = \frac{\omega \lambda_{ab}^2  k_s}{i c} \frac{1}{{\rm tanh} (k_s l_0 D)},
\end{eqnarray}
for the symmetrical, Eq.~\eqref{conts}, and anti-symmetrical, Eq.~\eqref{conta}, modes, respectively. The same expressions were derived in~\cite{plate} in the same geometry but without the effect of charge-neutrality breaking.

\subsection{Dispersion Relation}

Matching the ratios $(E_x/H_y)$ on the plate interface, Eq.~\eqref{ratio}, \eqref{ratios}, and~\eqref{ratioa}, and expressing hyperbolic cosines via $\kappa$, see Eq.~\eqref{kappa}, we obtain the dispersion relations for the eigenwaves in the following form,
\begin{equation}
\label{disp}
\frac{\sqrt{c^2k_x^2-\omega^2}}{r \omega^2/\omega_J}=1-\frac{\varkappa^{1-l_0}\pm \varkappa^{ l_0-1}}{\varkappa^{-l_0}\pm \varkappa^{l_0}},
\end{equation}
where symbols ``$+$'' and ``$-$'' stand for the symmetrical and anti-symmetrical modes, respectively,  $r=\lambda_{ab}^2\omega_J/c D$,
\begin{equation}
\varkappa=1+2\kappa-2\sqrt{\kappa+\kappa^2}.
\end{equation}

It should be emphasized that Eq.~\eqref{ks} has two different solutions since it is quadratic with respect to $\kappa$. Hence, there are two types of dispersion curves for each type of symmetry, see the following Section for details.

\section{Results and discussion}

\begin{figure}[t]
\centerline{\includegraphics[width=0.47\textwidth]{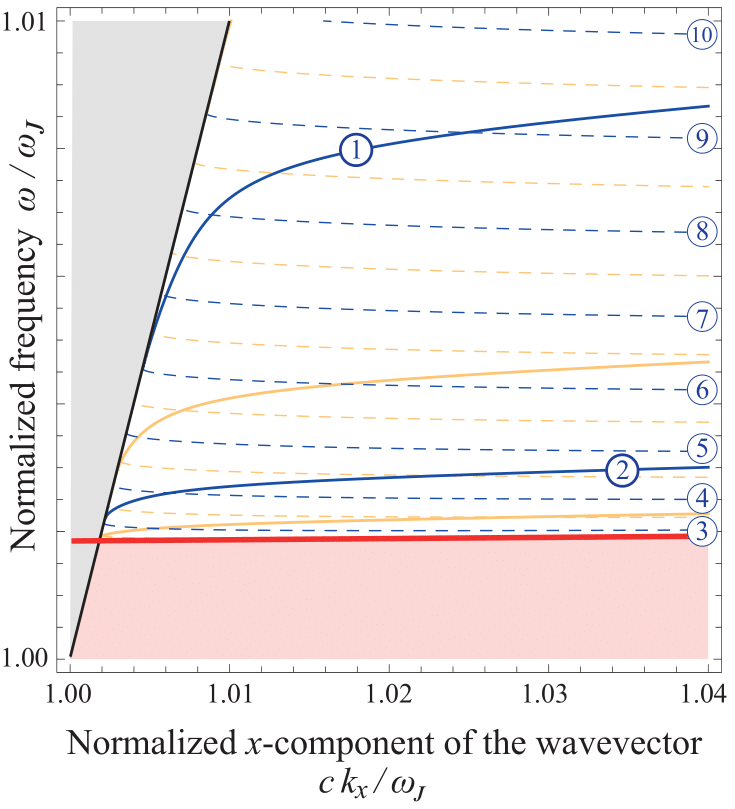}}
\caption{Localized waves spectrum for the high frequencies. Orange and blue solid lines are for the symmetrical and anti-symmetrical eigenmodes, respectively, of the 1-st type, and orange and blue dashed lines are for the eigenmodes of the 2-nd type. The curves for anti-symmetrical waves are numbered in ascending order of the wave number $k_s$. Black straight solid line is the light line $\omega=c k_x$. There is an additional forbidden region under the red line. Parameters: $\alpha=0.1,\, l_0=80,\, r=0.7,\, \gamma=\lambda_c/\lambda_{ab}=15,\, \varepsilon=16$.}
\label{fig2}
\end{figure}

\begin{figure}[t]
\centerline{\includegraphics[width=0.467\textwidth]{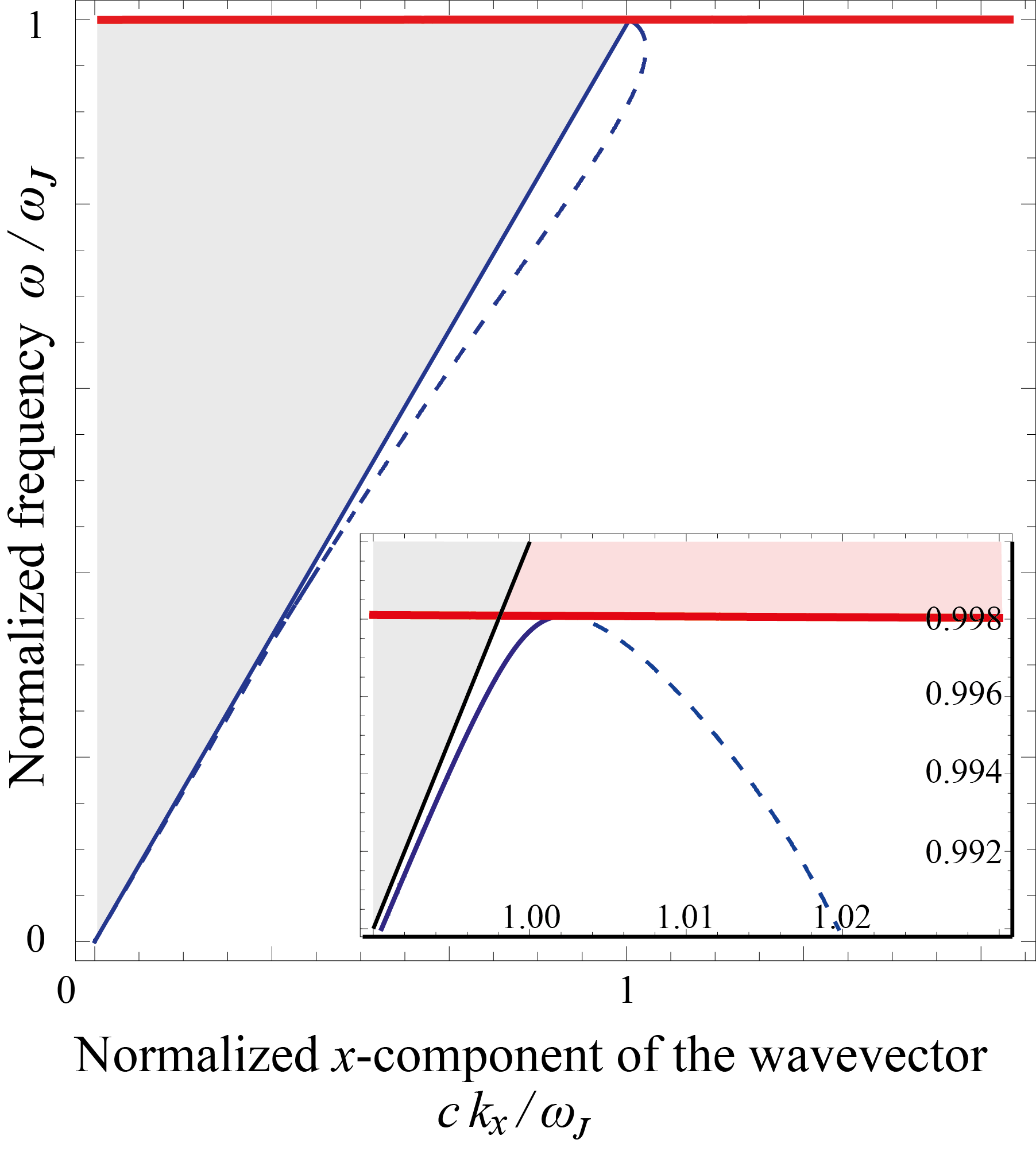}}
\caption{Localized (surface) waves spectrum for the low frequencies. Solid and
dashed blue lines are for the surface modes of 1-st and 2-nd types, 
respectively. The Inset panel shows the zoomed region where the dispersion 
curves join. The parameters are the same as in Fig.~\ref{fig2}.}
\label{fig3}
\end{figure}

The spectrum of localized eigenmodes determined by the dispersion relation~\eqref{disp} is plotted in Figs.~\ref{fig2} and \ref{fig3} for the characteristic values of parameters of high-temperature layered superconductors, such as $\rm Bi_2 Sr_2 Ca Cu_2 O_{8+\delta}$ or $\rm La_{2-x} Sr_{x} Cu O_4$~\cite{Laplace2016}. The non-zero coefficient $\alpha$ that characterizes the charge-neutrality breaking effect in the layered superconductors significantly affects the spectrum of localized waves.
Detailed analysis of the spectrum of the localized waves for the same geometry as in Fig.~\ref{fig1} but neglecting the effect of breaking of charge-neutrality, i.e., for $\alpha=0$, can be found in~\cite{plate}. In that case, there is one surface mode for $\omega<\omega_J$ and a set of waveguide modes with normal dispersion at $\omega>\omega_J$. Below we describe the effect of the charge-neutrality breaking, $\alpha>0$, on the spectrum, and indicate the main differences with the case of $\alpha=0$.

\subsection{Spectrum degeneration}

In Fig.~\ref{fig2}, one can see that there are two types of dispersion curves corresponding to two different solutions of Eq.~\eqref{ks} as a quadratic equation with respect to $\kappa$. Below we call these solutions as 1-st (``plus'') or 2-nd (``minus'') type with respect to the sign before the square root. 

In the continuum limit \eqref{conts}-\eqref{conta},  the dispersion equation \eqref{disp} with the 1-st type solution of \eqref{ks} transforms at $\alpha\to 0$ into the same equation as derived in paper \cite{plate} for not high frequencies, i.e., for $\omega\ll\gamma\omega_J$. Here $\gamma=\lambda_c/\lambda_{ab}$ is the anisotropy parameter of the layered superconductor, which is of order of $10\div100$ depending on material used. So, at small parameter $\alpha$, these curves (the solid lines in Fig.~\ref{fig2}) look similar to the ones reported in~\cite{plate}. 

The 2-nd type solution of \eqref{ks} gives a new, entirely different, dispersion equation resulting in a new set of curves shown by dashed lines in Fig.~\ref{fig2}. 

One can notice that the curves of the 1-st and the 2-nd type intersect, i.e., the spectrum is degenerated at certain parameters. These intersection points correspond to two different values of the wave number $k_s$ for different curves and the value of $k_s$ is higher for the dashed curves. The problem of the wave excitation and the question about which one of the two degenerated eigenmodes will be excited should be solved taking into account the so-called additional boundary conditions (see, e.g., \cite{bul}). 

As for the surface modes, they are practically the same for both symmetrical and anti-symmetrical fields (see Fig.~\ref{fig3}), and there is no intersection points in this frequency region $\omega~<~\omega_J$, i.e. no spectrum degeneration. Since $k_s^2<0$ in this frequency range, the spectrum of modes localized on the plate of finite thickness is similar to those for the modes localized on the half-infinite plate studied in~\cite{en}.

\subsection{Band gap}

In Figs.~\ref{fig2} and \ref{fig3}, one can see the light-red narrow forbidden region near the Josephson frequency. This region appears due to the condition \eqref{cond}. It is easy to see from this condition that the band gap increases with the growth of parameter $\alpha$ and the wave number $k_x$. The gap width is asymptotically proportional to $\sqrt{\alpha} k_x$ for the small parameter $\alpha$.

\begin{figure}[t]
\centerline{\includegraphics[width=0.47\textwidth]{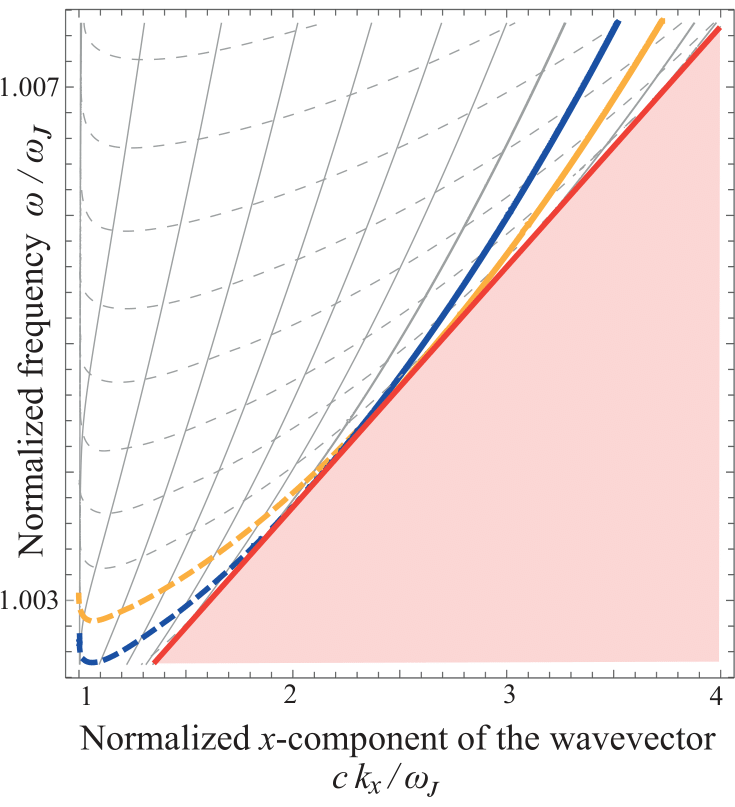}}
\caption{Joining of the high-frequency dispersion curves. Solid and dashed lines are for the 1-st and 2-nd type waves, respectively. The dispersion curves for one entire symmetrical and one entire anti-symmetrical modes are highlighted in color (orange and blue, respectively). The parameters are the same as in Fig.~\ref{fig3}}
\label{fig4}
\end{figure}

Note that the boundary of this region (shown by the red line) corresponds to the zero discriminant of \eqref{ks}. Therefore, the curves of the 1-st (solid) and the 2-nd (dashed) type are joined on this line. For the small wave numbers near the light line, this joining can be observed only for the surface mode (see inset in Fig. \ref{fig3}), while each of dashed (1-st type) curves in Fig.~\ref{fig2} joins with its solid (2-nd type) ``couple'' at higher wave-numbers $k_x$ (see Fig.~\ref{fig4}) forming one entire curve. 

\subsection{Anomalous dispersion}

Another interesting outcome of the charge-neutrality breaking is the appearance of the eigenmodes with anomalous dispersion.

All the dashed curves in Figs.~\ref{fig2},  \ref{fig3}, and~\ref{fig4} include sections with negative derivative $\partial \omega/\partial k_x < 0$. For the surface mode, the dispersion is normal right near the light line and then it becomes anomalous for a narrow wave-number range (see inset panel in  Fig.~\ref{fig3}). The waveguide modes start with the anomalous dispersion (if they include one) and then, at higher wave numbers, become normal (see Fig.~\ref{fig4}).

The anomalous dispersion can result in absolute wave instability under the propagation of a cylindrical beam of charged particles along the plate. This allows us to propose layered superconductors as the delaying media in {\it nanoscale} devices for the generation of surface and waveguide waves without a need to provide additional feedback in the system. 

\section{Conclusions}

In this theoretical paper, we investigate analytically the spectrum of localized JPWs in the layered superconductor with the effect of charge-neutrality breaking. We demonstrate that the non-zero coefficient characterizing this effect significantly affects the spectrum. First, the spectrum degenerates at frequencies higher than Josephson frequency: the dispersion curves of the first type, similar to the ones without charge-neutrality breaking, intersect with the new additional curves resulting from the effect. Second, near the Josephson frequency, the spectrum has a band gap whose width is proportional to the square root of the charge-neutrality breaking coefficient. The dispersion curves of both types are joining at the boundary of this forbidden region. Third, probably the most interesting in practical terms, all the additional dispersion curves include sections with anomalous dispersion near the light line. Therefore, layered superconductors can be used as a delaying media in nanoscale devices.

\section*{Acknowledgment}

N.K. gratefully acknowledges the support from the Austrian Science Fund (FWF), Grant I4865-N.

\end{document}